\documentclass[10pt,twocolumn,aps,superscriptaddress,pre]{revtex4-1}

\usepackage{amsmath}
\usepackage{graphicx}
\usepackage{subfigure}
\usepackage{epsfig}

\newcommand{\no}{\nonumber}

\begin{document}
\title{Travelling wave solutions to nonlinear Schrodinger equation with self-steepening and self-frequency shift}

\author{Anshul Saini}
\affiliation{Indian Institute of Science Education and Research (IISER)- Kolkata,
 Mohanpur, Nadia-741252, India}

\author{Vivek M. Vyas}
\affiliation{Indian Institute of Science Education and Research (IISER)- Kolkata,
 Mohanpur, Nadia-741252, India}
\email{vivek@iiserkol.ac.in}

\author{S. N. Pandey}
\affiliation{Motilal Nehru National Institute of Technology, Allahabad - 211004, India}

\author{T. Soloman Raju}
\affiliation{TempleCity Institute of Technology \& Engineering, Bhubaneswar 752 057, India}

\author{Prasanta K. Panigrahi}
\affiliation{Indian Institute of Science Education and Research (IISER)- Kolkata,
 Mohanpur, Nadia-741252, India}
\email{pprasanta@iiserkol.ac.in}

\begin{abstract}
We investigate exact travelling wave solutions of higher order nonlinear Schr\"odinger equation in the absence of third order dispersion, which exhibit non-trivial self phase modulation. It is shown that, the corresponding dynamical equation, governing the evolution of intensity in the femtosecond regime, is that of non-linear Schr\"odinger equation with a source. The exact localized  solutions to this system can have both super and subluminal propagation belonging to two distinct class. A number of these solitons exhibit chirality, thereby showing preferential propagation behavior determined by group velocity dispersion. Both localized bright and dark solitons are found in complementary velocity and experimental parameter domains, which can exist for anomalous and normal dispersion regimes. 
It is found that, dark solitons in this system propagate with non-zero velocity, unlike their counterpart in nanosecond regime. Interestingly, subluminal propagation is observed for solitons having a nontrivial $\text{Pad}\acute{e}$ type intensity profile.  
\end{abstract}

\maketitle

\section{Introduction}
Optical solitons were first predicted by Hasegawa and Tappert \cite{hasegawa} and later observed by Mollenauer {\em{et. al.}} \cite{mollen}. The existence of soliton solutons in non-linear Schr\"odinger equation (NLSE) is related to the complete integrability of this dynamical system \cite{zs}. It was found that non-linear Schr\"odinger equation aptly describes the pulse propagation in picosecond regime. Generalization of the NLSE is neccessitated to take into account higher order dispersion, self-steepening of the pulse due to the dependence of the slowly varying part of the non-linear polarization on time and the delayed effect of Raman response for describing optical pulse propagation in the femtosecond domain \cite{kh}. Higher order NLSE which captures these effects is given by \cite{potasek, agrawal},
\begin{align} \no
& i\frac{\partial \phi}{\partial z}-\frac{{\beta}_{2}}{2}\frac{\partial^{2}\phi}{\partial {\tau}^{2}}+{b}_{1}{\lvert\phi\lvert}^{2}\phi -\frac{i {\beta}_{3}}{6}\frac{{\partial}^{3}\phi}{\partial {\tau}^{3}}+i{b}_{2}\frac{\partial}{\partial \tau}({{\lvert\phi\lvert}^{2}\phi}) \\ & +i({b}_{3}+i{b}_{4})\phi\frac{\partial}{\partial \tau}({\lvert\phi\lvert}^{2})=0.
\end{align}
It has both group velocity dispersion (GVD) and third order dispersion term, respectively represented by $\beta_2$ and $\beta_3$. Parameter $b_1$ takes into account Kerr non-linearity, whereas self steepening effect arising from time variation of the non-linearity ${({\mid\phi\mid}^{2}\phi)}_{t}$ has been incorporated through the coefficient $b_2$. The terms with coefficient $b_3$ and $b_4$ account for the self frequency shift, originating from delayed Raman response.

This model, unlike NLSE, is not integrable in general. A few integrable cases have been identified; these are known as (i) Sasa-Satsuma case \cite{ss}, (ii) Hirota case \cite{hirota} and (iii) derivative NLSE of type I and type II \cite{ac}. A few restrictive special solutions of bright and dark type have been obtained for this system \cite{palacios,li,cnk}. Effects of these higher order terms on pulse propagation have been studied numerically quite extensively \cite{agrawal,band}; some special solutions to this system are also known \cite{olivra}.

Third order dispersion becomes important for femtosecond pulses, when group velocity dispersion (GVD) is close to zero. It can be neglected for optical pulses, whose width is of the order of $100$ femtoseconds or more, having power of the order of 1 Watt and GVD far away from zero. For short range propagation, $b_4$ can also be ignored. The resulting equation can be written in a dimensionless form by appropriately scaling variables:
\begin{align}
\psi = \mu \left( \frac{{b}_{1}}{\lvert{\beta}_{2}\lvert} \right)^{2} \phi, \, t=\frac{\tau}{\mu }, \,\,
\text{and}\,  x=\frac{\lvert{\beta}_{2}\lvert z}{{\mu}^{2}};
\end{align}
where $\mu=[\frac{1}{e}\times \text{half width}]$ of the incident pulse. The dimensionless equation obtained reads:
\begin{align}\no \label{HNLSE}
i\frac{\partial \psi}{\partial x}+{a_1}  \frac{\partial^{2}\psi}{\partial t^{2}} & + {\lvert\psi\lvert}^{2}\psi \\ & +i({a_4} \frac{\partial}{\partial t}({{\lvert\psi\lvert}^{2}\psi})+{a_5} \psi\frac{\partial}{\partial t}({\lvert\psi\lvert}^{2}))=0,
\end{align}
where,
\begin{eqnarray}
a_1=-\frac{sgn(\beta_2)}{2},
a_4=\frac{b_2}{b_1\mu}
 \ \text{and} \ \ a_5=\frac{b_3}{b_1\mu}.
\end{eqnarray}



In a previous study by some of the present authors, localized wave packets with nontrivial chirping were emerged as exact solutions of this equation \cite{pkp3}. It was found that chirping has a kinematic component, determined through initial conditions and a dynamical component, originating from self-steepening of pulse and its delayed Raman response. These two chirpings have inverse characteristics and can be used for chirp control. Further, periodic solutions, with sinusoidal functions in fractional linear form, have also been identified.

In this paper, we systematically investigate wide class of solutions of this system containing both bright and dark solitons. Interestingly, the equation governing dynamics of intensity is found to be the one that governs dynamics of field amplitude in non-linear Schr\"odinger equation with a source \cite{pkp1,pkp2}. This connection is then profitably used to obtain exact solitonic intensity waves of this higher order NLSE, possessing nontrivial chirping behavior. The NLSE type bright solitons are found to propagate at superluminal velocity, with dark solitons complementing them in parameter domains, which are experimentally accessible. Interestingly, both the solutions also complement each other in velocity space. Further, we find that dark soliton which are static in systems like NLSE, are propagating in this system solely due to presence of higher order nonlinearities. These solutions show chirality, which depends on the sign of GVD parameter. Depending on the kinematic chirp parameter two distinct domains are obtained. In the first case, if solitons are chiral for anomalous dispersion ($a_1>0$), then they are non-chiral for normal dispersion ($a_1<0$). In the second case, solitons are strictly chiral in both the dispersion domains. We then obtain more general localized solutions, through a conformal transformation, which can travel at both sub and superluminal velocities. In order to classify and gain physical understanding of the obtained solutions, we then use a pseudo-potential picture, which shows that the solutions separate into bounded periodic oscillations around the well minima and unbounded motion, separated by the so called separatrix solution. The separatrix solution is identified and the connection of this system with hydrodynamical equations are pointed out. Interestingly, the localized kink type solitons of this system representing the separatrix can be mapped to the separatix solutions of the elliptic function equation, through the aforementioned conformal transformation.

The paper is organized as follows. In the following section, we explore the localized dark and bright soliton solutions after connecting the present dynamical system with the NLSE. Chirality of the solutions and the effect of chirping have been systematically investigated. In Sec. III, we find more general solutions, through a fractional linear transform and show they can be both sub and super luminal depending upon various parameters. Section IV deals with conclusion and future directions of work.

\section{Exact solutions to higher order non-linear Schr\"odinger equation (HNLSE)}

We seek a travelling wave packet type solution of the HNLSE, which can generally be written as:
\begin{equation}\label{ansatz}
\psi(x,t)=\rho(\xi)e^{i\chi(\xi)}.
\end{equation}
\noindent Here the travelling coordinate $\xi = \alpha (t- ux)$, and $\rho$ and $\chi$ are real functions of $\xi$, $\alpha$ is the scale parameter and $u=1/v$, where $v$ is group velocity of the ansatz solution. Substituting (\ref{ansatz}) in (\ref{HNLSE}), and equating real and imaginary parts yields two coupled equations:
\begin{align} \label{c1}
\nonumber - {\alpha u \rho'} + {2 {\alpha}^2 {a_1} {\chi}' {\rho}'} & + {{\alpha}^2 {a_1} {\chi}'' {\rho}}  
\\ & + { 3 {\alpha} {a_4} {\rho}^2 {\rho}'} + {2 {\alpha}{a_5} {\rho^2} {\rho}'}=0,\\
\,\, \nonumber {{\alpha u} {\chi}' \rho} + {{\alpha}^2 {a_1}{\rho}''} & - {{{\alpha}^2 {a_1}}  {{\chi}'}^2 \rho} \\ &+ {a_2 {\rho}^3} - {\alpha}{a_4}{\chi}'{\rho}^3 = 0.\label{c2}
\end{align}
Equation (\ref{c1}) can be exactly integrated to yield:
\begin{equation}
\chi' = \frac{u}{2 \alpha a_1} + \frac{c}{\alpha a_1 \rho^2} - \frac{(3 a_4 + 2 a_5)}{4 \alpha a_1} {\rho}^2,\label{phase}
\end{equation}
where $c$ is a constant of integration, to be determined by initial conditions. Notice that the phase has a nontrivial form and has two intensity dependent chirping terms, apart from the kinematic first term, which is of usual $e^{i(kx-wt)}$ type. As is evident, the second term is also of kinematic origin and is infact present in the linear Schr\"odinger equation. The last term is due to higher order nonlinearities, which has a dynamical origin, and leads to chirping that is exactly inverse to that of the former. This is a novel form of self-phase modulation.

Using the above expression, Eq. (\ref{c2}) can be written as:
\begin{equation}\label{thetaeqn}
{\theta_1} \rho'' + {\theta_2} {\rho} + {\theta_3}{\rho}^3 + {\theta_4}{\rho}^5 = \frac{c^2}{\rho^3},
\end{equation}
\noindent with $\theta_1={\alpha}^2 {a_1}^2$, $\theta_2=\frac{({u^2}+{2 a_4 c}+{4 a_5 c})}{4}$,
$\theta_3= \frac{(2 a_1 - u a_4)}{2}$ and $\theta_4=\frac{(a_4 - 2 a_5)(3 a_4 + 2 a_5)}{16}$.

Note that, in parameter regime, where $\theta_{4}=0$, the amplitude dynamics will be like the NLSE case, albeit with dressed parameters. Phase dynamics, however will not be the same as that of NLSE, and will depend on, whether $(a_4 - 2 a_5)=0$ or $(3 a_4 + 2 a_5)=0$. In the latter case only, it will be exactly like NLSE, devoid of dynamical chirping. As shown in Ref. \cite{pkp3}, localized solutions of the above equation, in these parameter domains, show directionality and hence are chiral, and the propagation direction for these solitons is decided by sign of $a_{4}$. Because, these solutions satisfy NLSE with dressed parameters, many of the parametric restrictions on solution space are relaxed, for example both dark and bright solitons are present in both anomalous and normal dispersion regimes. 

Defining $k= {\theta_{1}} {\rho'}^{2} + \theta_{2} {\rho}^{2} +  \frac{\theta_{3}}{2} {\rho}^{4} + \frac{\theta_{4}}{3} {\rho}^{6} + \frac{c^{2}}{ {\rho}^{2}}$, Eq. (\ref{thetaeqn}) can be written conveniently in terms of $\sigma$ (where $\sigma={\rho}^{2}= \psi^{\ast} \psi$) to yield:
\begin{equation}
{\frac{\theta_1}{2} \sigma''} + 2 \theta_2 \sigma + {\frac{3 \theta}{2}}{\sigma^2} + {\frac{4 \theta_4}{3}}{\sigma}^3 = k,\label{sigma}
\end{equation}
where, $k$ is to determined via initial conditions. Motivated by the similarity of the above intensity equation with that of amplitude equation of NLSE in the presence of a source \cite{pkp1,pkp2}, below we consider different ansatz solutions.

\noindent {\bf{Case I}}: First we study localized soliton ansatz of the type
\begin{equation}
\sigma={\sigma_0} \left( 1+\gamma \,\mathrm{sech} (\frac{\gamma\xi}{\eta}) \right),
\end{equation}
which allows for both dark and bright solitons, depending on the sign of $\gamma$. Consistency of the above solution yields the following conditions:
\begin{align}
&-\frac{\theta_1\sigma_0}{{\eta}^{2}}+\frac{4{\theta_4}{\sigma_0}^{3}}{3}=0,\\
&\frac{3\theta_3{\sigma_0}^{2}}{2}+4{\theta_4}{\sigma_0}^{3}=0,\\
&\frac{{\theta_1}{\sigma_0}{\gamma}^{2}}{2{\eta}^{2}}+2{\theta_2}{\sigma_0}+3{\theta_3}{\sigma_0}^{2}+4{\theta_4}{\sigma_0}^{3}=0,  \\ \label{cc4}
\text{and} \,\,\,  &2{\theta_2}{\sigma_0}+\frac{3{\theta_3}{\sigma_0}^{2}}{2}+\frac{4{\theta_4}{\sigma_0}^{3}}{3}-k=0.
\end{align}
Solving first three of these relations, one finds
\begin{align}
& \sigma_{0}=-\frac{3\theta_3}{8\theta_4}, {\eta}^{2} =\frac{16\theta_1\theta_4}{3{\theta_3}^{2}}\\ 
\text{and} \,\,\, &{\gamma}^{2}=\frac{2}{3}\frac{(9{\theta_3}^{2}-32\theta_2\theta_4)}{{\theta_3}^2}.
\end{align}
Since, $\theta_1\ge0$, reality of $\eta$ demands $\theta_4\ge0$. Similarly, reality of $\gamma$ requires,  $9{\theta_3}^{2}-32{\theta_2}{\theta_4} > 0$, whereas positivity of $\sigma_0$ leads to $\theta_3 < 0$. These conditions, together with the above relations, lead to:
\begin{align}
(a_4 - 2 a_5) > 0,\,\, & u > \frac{2a_1}{a_4}, \, \text{and}\\ \no
\left(9{a_4}^{2}-2b \right) {u}^{2} - &\left(36 {a_1} {a_4} \right) {u} \\  +  &\left( 36 {a_1}^{2}- 4 b c({a_4}+2{a_5}) \right) > 0;
\end{align}
where $b=(a_4 - 2 a_5)(3 a_4 + 2 a_5)$.

In case of grey solitons, \emph{i.e.}, $\gamma<0$, for positivity of $\sigma$ requires $\gamma \geq -1$, which gives,
\begin{align}\no \label{dark}
\left(15{a_4}^{2}-2b \right) {u}^{2}- &\left(60 {a_1} {a_4} \right) {u}  \\ & +  \left( 60 {a_1}^{2}-8 b c({a_4}+2{a_5}) \right) \leq 0.
\end{align}
We note that $a_1$ can be positive or negative. Depending on the value of $a_4$, $ u > \frac{2a_1}{a_4}$ shows possible chirality of the solutions. In case when $\gamma = -1$, intensity vanishes at origin and these solutions are known as dark solitons. Interestingly, derivative of phase (from Eq.(\ref{phase})) becomes singular at that point, provided $c\neq 0$. This inturn implies that phase exhibits a discontinuous jumps across origin. Dark solitons of this kind have long been know in case of NLSE where they are strictly static \cite{zs}. Velocity space of dark solitons in our case can be investigated using Eq. (\ref{dark}). One can easily check that these solitons in general are not static, unlike NLSE, and exhibit superluminal propagation with velocity depending on $c$. However, if $(3 a_{4}+4 a_{5})(a_{4}+4 a_{5}) = 0$, then these solitons can be static. Interestingly, in the limit $a_{4,5} \rightarrow 0$ one can immediately see that Eq.(\ref{dark}) only allows for static solitons which corroborates with known results. This essentially implies that presence of higher order nonlinearities allows these kind of dark solitons to propagate with non-zero velocity.   
 


The last consistency relation (Eq.(\ref{cc4})) determines $u$, the soliton velocity in terms of $k$: 
\begin{equation}\label{u}
9{\theta_3}^3-48\theta_2\theta_3\theta_4-64{\theta_4}^2 k=0;
\end{equation}
we note that $\theta_2=\frac{({u^2}+{2 a_4 c}+{4 a_5 c})}{4}$ and $\theta_3= \frac{(2 a_1 - u a_4)}{2}$ contain $u$ explicitly. As defined earlier,
\begin{equation}
k= {\theta_{1}} \frac{{\sigma'}^{2}}{4\sigma} +{\theta_{2}}{\sigma} +  \frac{\theta_{3}}{2} {\sigma}^{2} +\frac{\theta_4}{3}{\sigma}^{3}
+ \frac{c^{2}}{\sigma},
\end{equation}
is a constant of motion, and hence its value remains unchanged under time evolution. At $\xi \rightarrow \infty$,  it is seen that $\sigma'\to 0$ and $\sigma \to \sigma_0$, which allows one to write:
\begin{equation}
k=-\frac{3}{8}\frac{\theta_2\theta_3}{\theta_4}+\frac{27}{512}\frac{{\theta_3}^{3}}{{\theta_4}^{2}}-\frac{8}{3}\frac{\theta_4 {c}^{2}}{\theta_3}.
\end{equation}
Using this, along with Eq. (\ref{u}), yields an equation for $u$ involving the chirp parameter:
\begin {equation}
135{\theta_3}^{4}-36\theta_2{\theta_3}^{2}b+{b}^{3}{c}^{2}=0.
\end {equation}
When the constant of integration, $c=0$ one finds a quadratic equation for u:
\begin{equation}\label{qu}
{u}^{2}(15{a_4}^{2}-4b)-60a_1a_2a_4u+60{a_1}^{2}=0.
\end{equation}
Reality of $u$ can be ensured by demanding positivity of the discriminant in the above equation,  which leads to $\theta_4\ge0$; this condition is trivially satisfied as it emerges from the reality of $\eta$.

We now systematically investigate the chirality of the solutions. For anomalous dispersion regime, $a_{1} > 0$, velocity is found to be positive and bounded from above, $0< v < \frac{a_{4}}{2 a_{1} }$. For $a_{1} < 0$,  $v \ge 0$ or $v < - \frac{a_{4}}{2 |a_{1}| }$. For $\theta_4 > 0$, it is straightforward to see that both the roots of Eq. (\ref{qu}) are negative, hence the possibility of $v \ge 0$ is forbidden. Hence, the solution is unidirectional and show chirality, with the direction of propagation governed by the sign of $a_{1}$. Typical experimental values corresponding to pulse propagation around $100$ femtosecond region are $a_1=0.5, a_4=10^{-1.5}, a_5=10^{-2}$. Using these values, one finds $v=0.01275$, which after appropriate rescaling leads to  $v=6.3\times10^{10}\,  \text{m/s}$, yielding superluminal velocity. We have checked that in the currently experimentally accessible parameter domain, all the solution are superluminal.

\noindent {\bf{Case II}}: Next we consider kink type solutions: 
\begin{equation}
\sigma={\sigma_0}\left( 1+ \gamma \text{tanh}(\frac{\gamma\xi}{\eta}) \right).
\end{equation}
where $\mid\gamma\mid<1$ due to positivity of intensity. Proceeding like the earlier case, one finds four consistency conditions, three of which yield:
\begin{eqnarray} 
\sigma_{0}&=&-\frac{3\theta_3}{8\theta_4},\\
{\eta}^{2} &=&-\frac{16\theta_1\theta_4}{3{\theta_3}^{2}},\\
\text{and} \,\, {\gamma}^{2}&=&\frac{1}{3}\frac{(9{\theta_3}^{2}-32\theta_2\theta_4)}{{\theta_3}^2}.
\end{eqnarray}
Reality of $\eta$ requires $\theta_4 < 0$, which along with positivity of $\sigma_0$ implies $\theta_3 >0$.
This clearly implies that, the kink type solution and the earlier found soliton live in mutually disjoint parameter and velocity spaces. Further, as in the earlier case, the reality of  $\gamma $ requires,
\begin{equation}
\left(9{a_4}^{2}-2b \right) {u}^{2}- \left(36 {a_1}{a_4} \right) {u} +  \left( 36 {a_1}^{2}-4 b c({a_4}+2{a_5}) \right) > 0, 
\end{equation}
to be satisfied. Since $\left(9{a_4}^{2}-2b \right) > 0$, the above inequality holds for 
all real values of $u$, provided the discriminant is negative. This leads to restrictions on the chirp parameter:
\begin{equation}
c\le-\frac{18{a_1}^{2}}{({a_4}+2{a_5})(9{a_4}^{2}-2b)}.
\end{equation}
When above relation is not obeyed, the allowed $u$ values are forbidden between the two roots.
As noted above, positivity of intensity demands $|\gamma|<1$, which translates to: 
\begin{equation}\label{tanhcon}
6{\theta_3}^{2}-32\theta_2\theta_4<0,
\end{equation}
imposing an upper bound on $c$:
\begin{equation}
c\le-\frac{6{a_1}^{2}}{({a_4}+2{a_5})(3{a_4}^{2}-b)}.
\end{equation}
Since, $\frac{6{a_1}^{2}}{({a_4}+2{a_5})(3{a_4}^{2}-b)}>0 $, for existence of these solutions, $c$ has to be a non-zero negative quantity. Like the earlier case, the consistency condition  $\theta_3 > 0$ requires
\begin{align}
\frac{1}{v} <\frac{2 a_{1}}{a_{4}}. 
\end{align}
For the normal dispersion region, this condition gives, $0 > v > -\frac{a_{4}}{2 \mid a_{1} \mid  a_{2}}$ and for anomalous region ($a_{1}>0$) above inequality leads to $v<0$ or $ v > \frac{a_{4}}{2 a_{1} a_{2}}$ , if
\begin{widetext}
\begin{equation}
-\frac{6{a_1}^{2}}{({a_4}+2{a_5})(3{a_4}^{2}-b)} \geq c \geq \frac{6{a_1}^{2}}{3{a_4}^{3}+2{a_4}^{2}{a_5}-12{a_5}^{2}{a_4}-8{a_5}^{3}}.
\end{equation}
\end{widetext}
When ${a_1}>0$, Eq. (\ref{tanhcon}) forbids the negative values of velocity. Hence for this interval, we have strict chirality for the solutions.

\begin{figure}[htp]
 \begin{center}
   \subfigure[$\Delta_{2}$ as a function of $f$ and $c$]{\label{fig:edge-a}\includegraphics[scale=0.5]{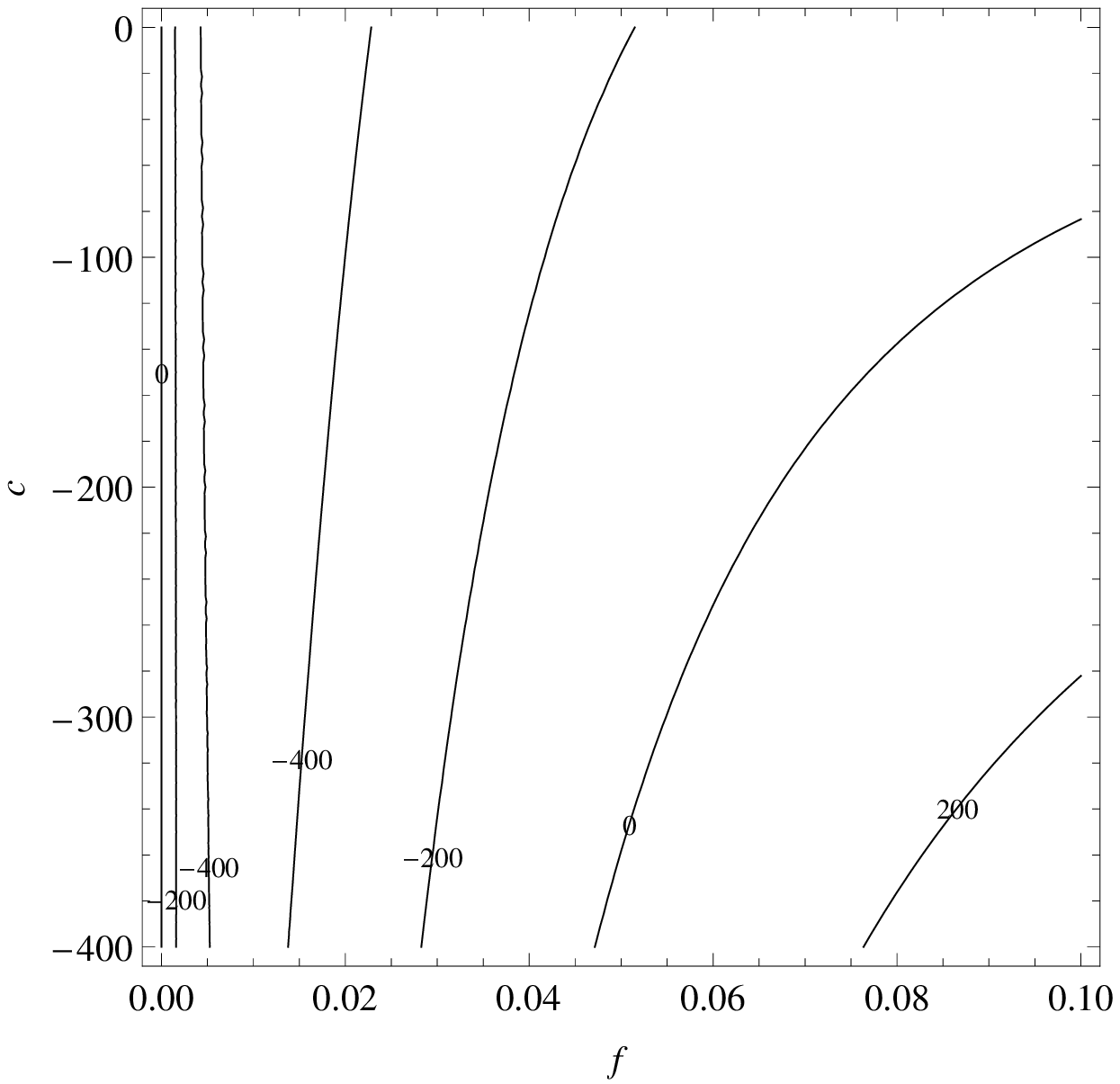}}
   \subfigure[$\Delta_{3}$ as a function of $f$ and $c$]{\label{fig:edge-b}\includegraphics[scale=0.5]{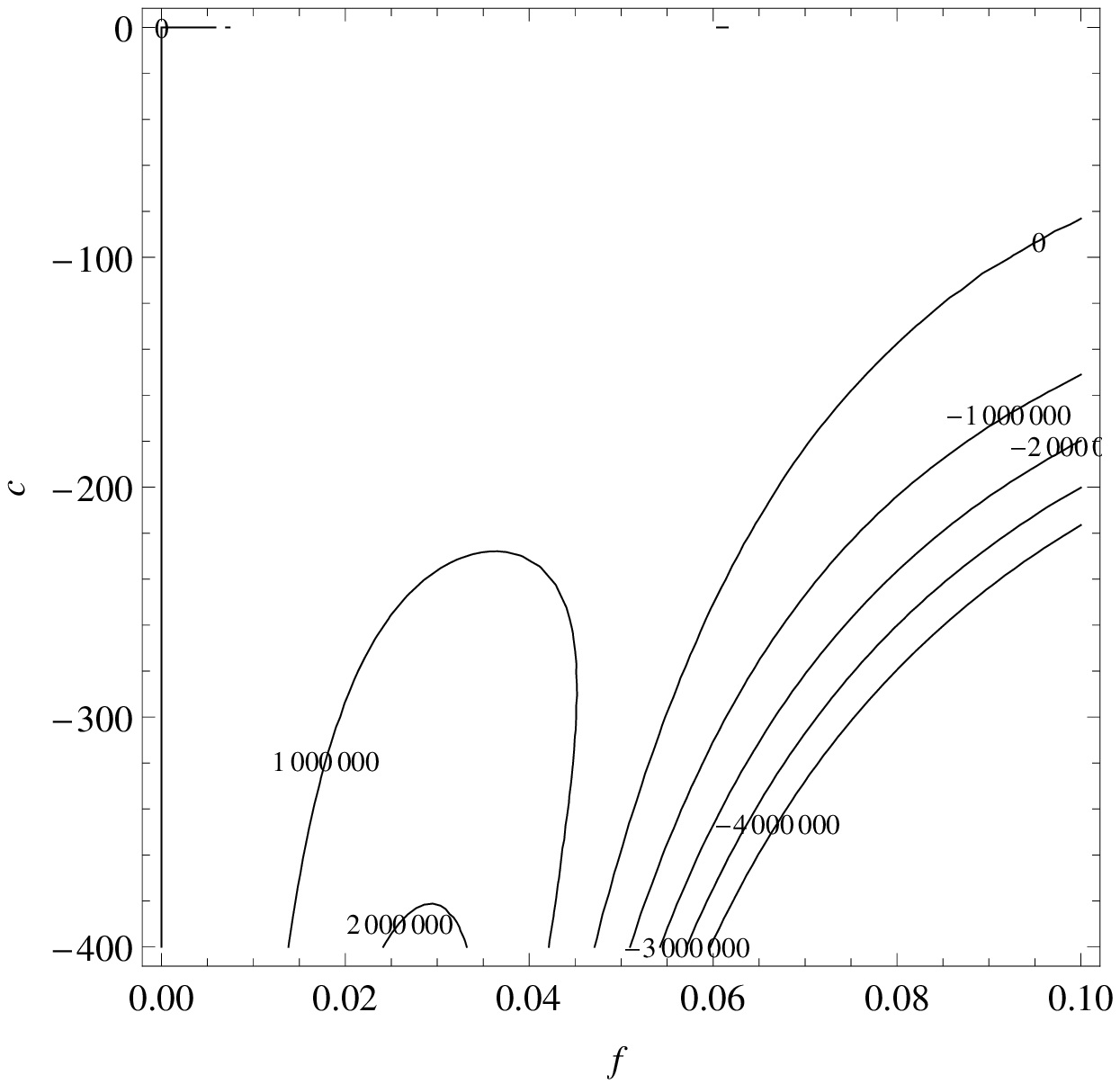}}
  \end{center}
\caption{$\Delta_{2,3}$ are plotted above for $a_1 = 0.5$, $a_4= 0.03162$, $a_5 = 0.01581+f$ and $c$ is chirping parameter}\label{fig}
\end{figure}

The remaining consistency condition yields an equation for $u$: 
\begin{equation}
9{\theta_3}^3-48\theta_2\theta_3\theta_4-64{\theta_4}^2 k=0,
\end{equation}
which has the same form as in the previous case, albeit the value of $k$ will be different. Constant of motion, $k$, is determined by evaluating it at $\xi = 0$ :
\begin{equation}
k=-\frac{3}{2}\frac{\theta_2\theta_3}{\theta_4}+\frac{21}{128}\frac{{\theta_3}^{3}}{{\theta_4}^{2}}-\frac{8}{3}\frac{\theta_4 {c}^{2}}{\theta_3}+\frac{2 {\theta_2}^{2}}{\theta_3}.
\end{equation}
Hence, equation for $u$ reads:
\begin {equation}
108{\theta_3}^{4}-72\theta_2{\theta_3}^{2}b+12{\theta_2}^{2}{b}^{2}+{b}^{3}{c}^{2}=0,
\end {equation}
which is a quartic equation. We have not been able to find real roots of this equation for experimental values around 100 femtosecond pulse width. This can be proved by looking at the sub determinant structure of this quartic equation, {\emph{i.e.}}, $\Delta_1, \ \Delta_2,\  \Delta_3,\ \text{and} \ \Delta_4 $ \cite{penn,quartic}. The sign of  $\Delta_i$ (where $i=1,2,3,4$), will determine the characteristics of the roots. Lets us take the list of signs of the $\Delta_i$, {\emph{i.e.}}, $[s_1,s_2,s_3,s_4]$, where $s_i$ is the sign of $\Delta_i$ . The number of sign change in the sequence reveals the number of real pairs. If the sequence has $n$ number of change in sign, then $n$ pair of complex roots exist. Here $\Delta_1,\ \Delta_4 >0$ are always satisfied in the allowed domain $i.e.$, $\theta_4 < 0 $. Hence for a real solution, we must have $\Delta_2, \Delta_3 >0 $. In Fig. \ref{fig} we have plotted $\Delta_2$ and $\Delta_3 $ as a function of $c$ and $f$ respectively, one can see from the graph that both quantity do not share same sign for any value of $c$ and $f$.

\section{Conformal transformation}

Apart from ansatz solutions discussed above, below we show that Eq.(\ref{sigma}) can be solved quite generally, using a conformal mapping involving Jacobi elliptic functions. 

\noindent {\bf{Case I}}:  When $\theta_{4} \neq 0$

In this case, Eq. (\ref{sigma}) can be written in a simpler form by going to $y$ variable, $y=\sigma + \frac{3 \theta_{3}}{8 \theta_{4}}$, 
\begin{equation}\label{nlsesource}
y'' + p y + q y^{3} + r = 0,
\end{equation}
where $p = \frac{2}{\theta_{1}}(2 \theta_{2} - \frac{9 {\theta_{3}}^{2}}{16 {\theta_{4}}^{2}}) $, $q = \frac{8 \theta_{4}}{3 \theta_{1}}$ and 
$r = \frac{2}{\theta_{1}} (\frac{9{\theta_{3}}^{3}}{64 {\theta_{4}}^{2}} - \frac{3 \theta_{2} \theta_{3}}{4 \theta_{4}} - k)$. In case when $r=0$, above equation coincides with the equation satisfied by Jacobi elliptic functions. Below we show that, for some real constants $A$, $B$ and $C$, the conformal transformation: 
\begin{equation}
y = \frac{A + B f}{1 + C f},\label{solun}
\end{equation}
solves equation (\ref{nlsesource}), with $f(\xi,m)$ being one of the twelve Jacobi elliptic functions with modulus parameter $m$ \cite{pkp1,pkp2}. These elliptic functions satisfy: 
\begin{equation}\label{elliptic}
f''= a f + b f^{3}, 
\end{equation}
with appropriate real constants $a$ and $b$. For example, for $f=cn(\xi,m)$, $a=(2m-1)$ and $b=-2m$. 
The claim that expression (\ref{solun}) solves Eq. (\ref{nlsesource}) can readily be seen, by defining the first integral $E_{0}=\frac{{f'}^{2}}{2} - \frac{af^{2}}{2} - \frac{bf^{4}}{4}$, and substituting (\ref{solun}) in Eq. (\ref{nlsesource}). Use of Eq. (\ref{elliptic}) leads to the following consistency conditions:
\begin{align}
& -4 B C E_{0}+ 4 A C^2 E_{0} + p A  + q A^3 + r  = 0,\\
& a B  - a A C + p B  + 2 p A C + 3 q A^2 B + 3 r C = 0,\\
& -a B C + a A C^2 + 2p B C + p A C^2 + 3 q A B^2 + 3 r C^2 = 0,\\
& b B  - b A C + p B C^2 + q B^3 + r C^3 = 0.
\end{align}
Notice that these coupled algebraic equations are nonlinear in $A$, $B$ and $C$, but are linear in $p$, $q$, $r$ and $E_{0}$, and hence can be solved exactly to yield: 
\begin{align}
& E_{0}=-\frac{b A +a B C+a A C^2}{4 B C^3},\\
& p=\frac{-3 b A  - 2 a B C - a A C^2}{C (-B +A C)},\\
& q=\frac{-b -a C^2}{B (B -A C)},\\
& r=-\frac{b A B  + b A^2 C + a B^2 C + a A B C^2}{C^2 (B -A C)}.
\end{align}
Therefore, this shows that, provided above relations hold, Eq. (\ref{nlsesource}) is indeed solved by (\ref{solun}). Notice the dependence of $E_{0}$ on $A$, $B$ and $C$, which simply shows that the initial values in Eq. (\ref{elliptic}) do play a role in determining $A$, $B$ and $C$, and hence they can not be fixed, given $p$, $q$ and $r$. Since, original equation parameters $a_{i}$ ($i=1,...,5$) can be expressed in terms of $p$, $q$ and $r$, one can determine the parameter regime for a given solution using these relations. The solutions presented in Ref. \cite{pkp3} form a subclass 
of the ones found above. For $m=1$, one gets localized solutions and are often of experimental and technological interest. Setting $m=0$, singles out periodic solutions, which were reported in Ref.\cite{pkp3}.

For explicitness, we consider $f=\text{sech}(\xi)$, which gives $a=1,b=-2$ and $E_0$ is found to be 0.
The consistency equation becomes
\begin{eqnarray}
-2A + B C + A {C}^{2}=0,
\ \ -\frac{2 A C + B } { B - A C}=p,\\
\frac{C}{A(B-AC)}=q,\ 
\ and \ \ \ \frac{2{A}^{2}}{C(B-AC)}=r\\
\end{eqnarray}
After eliminating $B$ and $C$ one gets,
\begin{equation}
r+pA+q{A}^{3}=0,
\end{equation}
implying that velocity of soliton is governed by background amplitude in this non-linear system. Since we know the values of $p$, $q$ and $r$, by fixing the value of $u$, we can get the corresponding values of $B$ and $C$ in the form,
\begin{align}
{C}^{2}=\frac{2{A}^{3}q}{r} \, \text{and} \,  {B}^{2}=\frac{2{p}^{2}A}{qr}-8{A}^{2}.
\end{align}
For the experimental values, $a_1=0.5$, $a_4 = 0.0316228$, $a_5 = 0.01$ and putting $u=10^5$ (sub-luminal), we get $\text{A}=1.009\times 10^{7}$,\ $\text{B}=-\  6.51\times10^{7}$ and $\text{C} = 6.7453$. In the case of super-luminal propagation, $u=10$, we get $\text{A}=1536.46, \text{B}=1307.78$ and $\text{C}=1.05128$. As is evident, $A,B \propto u$ and $C\sim O(1)$.

For $f=\text{tanh}(\xi)$, in Eq. (\ref{elliptic}) one finds $a=-2\ b=2$ and $E_0=0.5$. In this case, the consistency conditions become
\begin{align}
\frac{-A+BC+AC^{2}}{BC^{3}}&=1,\\
\frac{-6A+4BC+2AC^{2}}{C(-B+AC)}&=p,\\
\frac{-2+2C^{2}}{B(B-AC)}&=q,\,\, \text{and}\\
\frac{-2AB^{2}-2A^{2}C+2B^{2}C+2ABC^{2}}{C^{2}(B-AC)}&=r.
\end{align}
After simplification, we get a quadratic equation for $u$ along with other consistency relations given by,
\begin{align}
& u^2(\frac{9{a_{4}}^{2}}{4}+\frac{b}{2})-9 a_1 a_4 u +(a_4+2 a_5) b c + b{a_1}^2=0, \label{ueq} \\
& 9 {\theta_{3}}^{3} - 48 \theta_{2} \theta_{3} \theta_{4} - 64 {\theta_{4}}^{2} k =0, \label{Aeq}\\
& C^2=-\frac{4 \theta_{4}}{3 \theta_{1}}A^2, \label{Ceq}\\
& B=\frac{A}{C}.\label{Beq}
\end{align}
Once $u$ is known, $A$ can be evaluated from Eq. (\ref{Aeq}) through $k$ dependence on $A$, then $C$ and $B$ can be obtained using Eq. (\ref{Ceq}) and (\ref{Beq}) respectively. For the consistency of Eq. (\ref{Ceq}), $\theta_4 < 0$ must be satisfied. Regularity of solution requires  $C<1$. Also, the positivity of $\sigma$ demands $\frac{A-B}{1-C}-\frac{3 \theta_3}{8 \theta_4}>0$, which gives restriction on the velocity of the soliton.

Above obtained solutions can be better understood using following pseudo particle picture. Eq. (\ref{nlsesource}) can be thought of as equation of motion for a classical particle of unit mass, with displacement given by $y$, moving under influence of a nonlinear force. The same also holds for Eq. (\ref{elliptic}), albeit the nonlinear force does not have a constant term. Also notice that both these dynamical systems are of one degree of freedom and their phase spaces are two dimensional, only position $y$ and momentum $y'$ is required to describe the dynamics completely. Looked in this setting, mapping Eq. (\ref{solun}) simply says how the two phase spaces are related; more precisely it shows how phase space for Eq. (\ref{nlsesource}) can be generated by knowledge of phase space for Eq. (\ref{elliptic}). Since, the mapping is conformal, the singularity structure of both the phase spaces are identical, modulo movable poles. Infact, it indicates that Eq. (\ref{nlsesource}) is integrable in the sense of Painlev\'{e}, since Eq. (\ref{elliptic}) is integrable \cite{ablo,gr}. The kink type solutions of the parent Eq. (\ref{HNLSE}), in this setting, can be identified with the separatrix solutions of Eq. (\ref{elliptic}) via the conformal mapping \cite{zas}. Similarly, singular solutions of Eq. (\ref{elliptic}) can be seen to be related to unbounded solutions lying outside separatrix, and periodic solutions can be seen to be related to bounded solutions lying inside separatrix. 

It is very interesting to note that solutions to Eq. (\ref{nlsesource}) are actually solutions to modified KdV equation: $v_{xxx} + p_{1} v^{2} v_{x} + p_{2} v_{t}=0$. This can be seen using the travelling variable $\zeta=x-vt$, integrating out the equation once, and identifying $p_{1}=3q$, $p_{2}=-\frac{p}{v}$ and $r$ as the constant of integration.   
So, at the traveling variable level, which restricts one to only one soliton solution, one can say that the modified KdV solitons can be simulated by considering these higher nonlinearities in optical fiber.

\noindent {\bf{Case II}}: When $\theta_{4}=0$

In this case, the above analysis is not valid since the degree of the Eq. (\ref{thetaeqn}) changes and can be written as:
\begin{equation}\label{thetaeqn2}
 {\sigma}'' + p' \sigma + q' {\sigma}^{2} + r' = 0,
\end{equation}
with $p'=\frac{4 \theta_{4}}{\theta_{1}}$, $q'=\frac{3 \theta_{3}}{\theta_{1}}$ and $r'=-\frac{2 k}{\theta_{1}}$. The above equation can be mapped to Eq. (\ref{elliptic}) via a transformation:
\begin{equation}
 \sigma = \frac{A + B f^{2}}{C + D f^{2}},
\end{equation}
which is along the same lines as (\ref{solun}). Following the same procedure as the former case, using the first integral $E_{0}=\frac{{f'}^{2}}{2} - \frac{af^{2}}{2} - \frac{bf^{4}}{4}$, we find the consistency conditions as:
\begin{align}
& 4 B C^2 E_{0} - 4 A C D E_{0} + p' A C^2 + q' A^2 C + r' C^3 = 0,\\
\nonumber & 4 a B C^2 - 4 a A C D - 12 B C D E_{0} + 12 A D^2 E_{0} + p' B C^2 \\ & + 2 p' A C D + 2q' A B C + q' A^2 D
+3 r' C^2 D = 0,\\ \nonumber
& 3 b B C^2 - 3 b A C D - 4 a B C D + 4 a A D^2 + 2 p' B C D + p' A D^2 \\ & + q' B^2 C + 2 q' A B D + 3 r' C D^2 =0,\\
& -b' B C D + b' A D^2 + p' B D^2 + q'B^2 D + r' D^3 =0. 
\end{align}
Again we observe that, the above equations are linear in $p'$, $q'$, $r'$ and $E_{0}$, and hence can be solved to give:
\begin{widetext}
\begin{align}
E_{0} & = -\frac{b C^2-2 a C D}{4 D^2},\\
p' & = -\frac{2 (-3 b B C^2- 3 b A C D + 4 a B C D + 2 a A D^2)}{D(A D -B C )},\\
q' &= \frac{6 (-b C^2 + a C D)}{A D -B C},\\
r' &= -\frac{-b B^2 C^2 - 4 b A B C D + 2 a B^2 C D - b A^2 D^2 + 4 a A B D^2}{D^2 (B C-A D)}. 
\end{align}
\end{widetext}
This shows that, the Eq. (\ref{thetaeqn2}) is an integrable equation, in the sense of Painlev\'e, and the explicit solutions can be expressed in terms of Jacobi elliptic functions. By choice, of appropriate values of $A$, $B$, $C$ and $D$ one can find out the parameter regime in which the given solution is valid. Further, solutions to Eq. (\ref{thetaeqn2}) actually satisfy KdV equation: $v_{xxx} + p_{1} v v_{x} + p_{2} v_{t}=0$, in travelling variable $\xi=x-vt$, with $p'=-v \beta$, $2q'=\alpha$ and $r'$ being an integration constant. It is very interesting to note that the condition $\theta_{4}=0$, can be fulfilled if $a_{4}=2 a_{5}$ or $3 a_{4} = -2 a_{5}$. In the case when the former is true then, the intensity profile of these solutions is exactly like NLSE solutions except with a non-trivial phase chirping. In the latter case, the solutions do not have this non-trivial chirping and the solutions are indistinguishable from NLSE solutions with appropriate coefficients.  

\section{Conclusion}

In conclusion, we have shown that nonlinear Schr\"odinger equation in the presence of self-steepening and self-frequency shift, possesses rich travelling wave dynamics, with non-trivial chirping. Self-phase modulation revealed here indicates the possibility of effective control of pulse dynamics through chirp management. The dynamics of the intensity is governed by the non-linear Schr\"odinger equation with a source. Hence, intensity waves show solitonic behavior, in complete analogy with the integrable NLSE. These solutions exhibit superluminal propagation in experimentally accessible parameter domain. Further, these solutions reveal directional propagation, akin to hydrodynamical equations like KdV, with which the obtained solutions have close connection. We have found that dark solitons in this system can propagate with nonzero velocity, unlike in the case of NLSE. More general solutions obtained through fractional linear transform, showed both sub and superluminal behavior. We identified the separatrix solution bounding the regular oscillatory  motion from the unbounded ones. This strengthens the hope that this system may be integrable, since separatrix is not known to exist in a system with chaotic dynamics.

\end{document}